\documentclass[12pt]{article}
\usepackage{latexsym}%showkeys,amssymb

\renewcommand{\theequation}{\thesection.\arabic{equation}}
\csname @addtoreset\endcsname{equation}{section}

\begin{document}

\begin{titlepage}
\begin{flushright}
NYU-TH/00/09/10 \\
SNS-PH-00-14 \\
hep-th/0009239
\end{flushright}
\vfill
\begin{center}
{\LARGE\bf Covariant Quantization \\ of the Brink-Schwarz Superparticle}    \\
\vskip 27.mm  \large
{\bf  P.~A.~Grassi$^a$, G.~Policastro$^{a,b}$,  and  M.~Porrati$^a$}\\
\vfill
{\small \it
$(a)$Physics Department,  New York University, \\ 
4 Washington Place, New-York,  NY 10003,  USA \\ 
\vspace{.2cm}
$(b)$ Scuola Normale Superiore, \\
Piazza dei Cavalieri  7, Pisa, 56100, Italy \\}
\end{center}
\vfill

\begin{center}
{\bf ABSTRACT}
\end{center}
\begin{quote}
  The quantization of the Brink-Schwarz-Casalbuoni superparticle is
  performed in an explicitly covariant way using the antibracket
  formalism.  Since an infinite number of ghost fields are required,
  within a suitable off-shell twistor-like formalism, we are able to
  fix the gauge of each ghost sector without modifying the physical
  content of the theory. The computation reveals that the antibracket
  cohomology contains only the physical degrees of freedom.  \vfill
  \hrule width 5.cm \vskip 2.mm
\end{quote}
\begin{flushleft}
September 2000
\end{flushleft}
\end{titlepage}

\newcommand{\note}[1]{{\tiny (note)}\marginpar {\scriptsize #1}}
\renewcommand{\thefootnote}{\fnsymbol{footnote}}
\newcommand{\ghe}[2]{$\stackrel{\textstyle #1}{\scriptstyle (#2)}$}
\newcommand{\lre}{\multicolumn{3}{c}{$\stackrel{*}{\longleftrightarrow}$}} 

\def\SL#1{\rlap{\hbox{$\mskip 4.5 mu /$}}#1}     % slash{#1}
\renewcommand {\theequation}{\arabic{section}.\arabic{equation}}

\section{Introduction}
\label{sec:intro}

%%%%%%%%%%%%%%%%%%%%%%%%%%%%%%%%%%%%%%%%

The correct quantization of systems with an infinitely reducible gauge
symmetry is a longstanding problem and, recently, several new efforts
(see e.g. \cite{berko}) have been made toward the construction of a
covariant quantization procedure for the Green-Schwarz superstring
model \cite{green3}. Indeed, the GS superstring is the most important
and interesting model enjoying the feature of infinite reducibility,
but since the model is very difficult to handle in its complexity, the
study of simpler models provides a good test for the quantization
techniques. This is essentially the reason why, in the past decades,
people devoted several efforts trying to quantize the superparticle
model of Brink-Schwarz-Casalbuoni type \cite{Brink-Schwarz}.

In the case of GS superstring and superparticle, there are first- and
second-class constraints \cite{Dirac,henneaux,gitman,bht}.  The occurrence
of second-class constraints arises from the fact that the Grassman
momenta $P^\alpha_\theta$, conjugate to the fermionic variables
$\theta_\alpha$, are non-independent phase-space variables.  If, as a
formal procedure, one attempts to construct Dirac brackets, treating
all the fermionic constraints as if they were second class, the
resulting expressions are singular. An alternative procedure would be
a careful separation of first- and second-class constraints, but, in
that way, a covariant quantization procedure is impossible to achieve.

In order to maintain manifest Lorentz covariance, one has to exploit
the $\kappa$-symmetry \cite{green3,Siegel:1983hh} of the model, which
cancels half of the fermionic degrees of freedom realizing the
matching between the bosonic and the fermionic states.  Unfortunately,
the $\kappa$-symmetry is a reducible local fermionic symmetry. This
means that, in reality, only four degrees of freedom are effectively
canceled at the first stage. Pursuing the analysis, it is easy to show
that an infinite tower of ghosts is necessary to match the correct
number of degrees of freedom.
In terms of the Hamiltonian formalism \cite{gitman}, this is
equivalent to the statement that the constraints are infinitely
reducible: there exist not only linear vanishing combinations of
constraints, but also zero modes of those relations, and so on to
infinitely many levels.

Several attempts to a solution of the problem can be found in the
literature. In particular, we would like to mention the idea of
changing the classical constraints in order that all the second-class
constraints are transformed into first-class ones \cite{GH,ilk}. This
essentially yields an extension of the phase-space where the
$\kappa$-symmetry is gauged by means of suitable fermionic gauge
fields.  Other approaches to the superparticle quantization are the
harmonic superspace \cite{harmo} --which produces a non-local super
Yang-Mills field theory-- and the construction of models based on a
given BRST operator which selects the correct physical subspace
\cite{kallosh}.  Nevertheless, all of these approaches share the
common feature of infinite number of classical and ghost fields.

Finally, the computation of the gauge-fixed BRST characteristic
cohomology \cite{anti_coho,barnich} for the superparticle model
received considerable attention
\cite{kallosh,Lindstrom_BS,Lindstrom,Bastianelli}. It has been shown
in \cite{Bastianelli} that, due to the infinitely many interacting
fields, the canonical transformations performed to implement the gauge
fixing turn out to be ill-defined.  Moreover, as a consequence of
further gauge symmetries of the gauge-fixed action, a two-step
gauge-fixing procedure is required and it may cause problems, as
argued in \cite{kallosh}.

The aim of our paper is to present a new solution of the application
of the BV-BRST formalism (\cite{BRST,BV}; for a review, see
\cite{henneaux,cano,gomis}) to the
problem of the superparticle. In particular, we take the advantage of
the existing literature to make essential steps towards the complete
solution. The main issue here is the construction of a procedure to
quantize the Brink-Schwarz-Casalbuoni classical action, computing the
correct antibracket map and the corresponding cohomology. The present
technique is based on canonical transformations which implement the
gauge fixing of the $n^{\rm th}$ order ghosts without affecting the
$(n-1)^{\rm th}$ order.

The structure of the superparticle model involves word-line
diffeomorphisms, on-shell closure of the algebra of symmetries and an
infinite tower of ghost fields. To take into account the open algebra,
the BV-BRST formalism is implemented and the solution of the master
equations contains quadratic terms in the antifields. On the other
hand, the $\kappa$-symmetry of the model entails field-dependent
transformations which produce interactions among the ghost fields of
different levels and the superparticle fields. Therefore, after the
gauge fixing, an infinite number of ghosts fields interact, among
themselves and with the {\it physical} fields.  In this situation, two
types of problems arise: {\it i)} the renormalizability of the model
may be lost due to an infinite number of parameters, {\it ii)} the
computation of the gauge-fixed BRST cohomology and the definition of
the physical spectrum of the model are ill-defined.

In order to solve the problem of interactions among the ghost fields
of different levels, we apply the idea of P.~Townsend~\cite{3lectures}
for the chiral superfield to the superparticle model.  The basic idea
is a suitable redefinition of ghost fields, in such a way that they
appear decoupled from the rest of the theory. The complete formulation
of Townsend's idea has already been discussed in \cite{poli_1}. There,
it was shown that, in the case of the gauged, complex, linear
superfield model \cite{GPZ}, a convenient redefinition of ghost fields
realizes the decoupling and the quantization can be performed. In
addition, it was shown that a careful application of the technique of
\cite{GPZ} provides a quantization procedure with the correct
antibracket cohomology. The realization of Townsend's idea in the
superparticle framework requires further efforts.

Following the twistor-like formulation \cite{twistor-like}, we
introduce twistor variables to re-express the ghost fields in a
decoupled fashion. In the literature only the on-shell twistor-like
formalism is available.  Unfortunately, this is not suitable for our
purposes. Therefore, we extended the twistor-like formalism in order
that the ghost fields can be conveniently redefined off-shell.  In
past years, some interesting formulations of the superparticle model
have been discussed in the context of the $N=8$ word-line supergravity
framework \cite{twistor-like}. However, we do not follow these
directions and our {\it twisted}-ghost formalism only amounts to a
redefinition of classical variables.

Even after the ``twisting'', the zero modes of the $\kappa$-symmetry
(the ghost fields) turn out to interact with the physical fields;
however, the interaction terms are proportional to the einbein
equation of motion (Virasoro constraints) and, therefore, these
couplings can be eliminated by a simple canonical transformation (cf.
Ref.~\cite{cano}). Although the fermion generator of these
transformations contains a formal sum on infinite ghost fields, it can
be easily verified that each individual canonical transformation
involves a finite number of fields \cite{kallosh,Bastianelli} and it
has a well-defined inverse. In this way, the cohomology of the
antibracket is not changed and these transformations do not affect the
physical observables.
After twisting and canonical transformations, the ghost 
fields are free and decoupled from the rest of the fields. In
addition, the ghost fields are {\it off-shell} zero modes of the
$\kappa$-symmetry. Comparing with \cite{poli_1,GPZ}, we stress again
the fact that the quantization procedure, in the presence of an
infinite number of ghosts, can be performed only in the case
of off-shell decoupling.

Although the technique of \cite{GPZ} appears very promising for the
superparticle models, as it has been shown in \cite{poli_1}, it cannot
be easily implemented in the the present context. This is due to the
structure of the antifield-dependent part of the action. The
diagonalization of the fields necessary to achieve the decoupling
among the ghosts of different levels induces new couplings in the antifield
terms which cannot be removed. Therefore, in order to apply
safely the quantization procedure of \cite{poli_1,GPZ}, one has to
reach the third order of ghost field, at least, to circumvent the
problem.  Unfortunately, despite several efforts, it seems that there
is no way to apply the aforementioned procedure to the superparticle case.
However, we can follow a similar strategy: we quantize the $n^{\rm th}$
ghost field without affecting the $(n-1)^{\rm th}$ level, but the
actual realization is different.

We choose the conventional Lorentz invariant gauge fixing proposed in
\cite{GH,kallosh,Lindstrom_BS} to fix the gauge of the spinor field
$\theta_\alpha$ and for the ghost fields. In particular, following
\cite{kallosh,stu}, we take into account the St\"uckelberg symmetry
arising in the {\it non-minimal} sector. The latter contains
Lagrangian multipliers, anti-ghost fields and the non-minimal
extra-ghost fields. Contrary to \cite{kallosh,kallo}, we use the
St\"uckelberg symmetry at each step of the quantization procedure in
such a way that all degrees of freedom are gauge fixed and their
dynamics is precisely prescribed.

Along the quantization procedure, we compute the antibracket
cohomology and we check that
the sequence of canonical transformations does not modify the physical
sector of the Hilbert space. In addition, we compute the cohomology
group $H(\gamma,H(\delta))$ of the completely gauge-fixed theory
--where $\gamma$ and $\delta$ are defined in \cite{anti_coho,gomis}--
and the result coincides with the correct physical spectrum.

The outline of the paper is the following. In section \ref{sec:BV}, we
briefly review the antifield and the antibracket formalism with
particular attention to the definition of physical observables. In
section \ref{sec:strategy}, we present some details about the strategy
of the quantization procedure. In section \ref{sec:solution}, we recall
the solution of the master equation in order to establish our
conventions. Next, in section \ref{sec:twist} we introduce the
twistor-like variables and we define the off-shell zero modes.
Finally, in section \ref{sec:gauge} the gauge fixing of the complete
tower of ghost fields is implemented. Section \ref{sec:cohomology}
contains the computation of the cohomology and concluding remarks. An
appendix details our notations.

%%%%%%%%%%%%%%%%%%%%%%%%%%%%%%%%%%%%%%%%

\section{Symmetries and the anti-bracket formalism}
\label{sec:BV}

Essential ingredients of the Batalin-Vilkovisky (BV) formalism are
anti-fields and anti-brackets \cite{BV,cano,gomis}.  For any field
$\phi ^A$, one introduces an anti-field $\phi ^*_A$ with opposite
statistics.  A ghost number is assigned to every field, in such a way
that for the classical fields $gh(\phi) =0$, the (extended) action has
ghost number zero,
and for all fields $gh(\phi^*_A )= -gh(\phi ^A)-1$. \\

The {anti-bracket} between two functions $F$ and $G$ of
the fields and anti-fields is defined by
\begin{equation}\label{abracket}
\left( F, G \right) = 
\frac{\partial_r }{\partial\phi ^A}  F \frac{\partial_l }{\partial\phi
  ^*_A} G -  
\frac{\partial_r }{\partial\phi ^*_A} F  \frac{\partial_l
  }{\partial\phi ^A} G.   
\end{equation}
where $\partial_{r/l}$ denote the derivative from the left or from the
right.   
The antibrackets satisfy graded commutation, distribution, and Jacobi
relations \cite{BV}.  
With these brackets, fields and anti-fields behave as coordinates and
their canonically conjugate momenta   
\begin{equation}\label{eq:brackffs}
(\phi^A,\phi^B)=0\ ;\hspace{0.5cm}
(\phi^*_A,\phi^*_B)=0;\hspace{0.5cm}
(\phi^A,\phi^*_B)=\delta^A_B.
\end{equation}
The extended action $S(\phi,\phi^*)$ is a solution of the master equation  
\begin{equation}\label{master}
(S,S)=0\, ,
\end{equation}
which satisfies certain boundary conditions:
1) it coincides with the classical action $S_{\rm cl}(\phi)$
when the ghost fields are set to zero; 2) the
anti-fields $\phi^*_A$  
are coupled to the gauge transformations of the classical fields; 3)
it satisfies the \emph{properness} requirement, i.e. the $2N \times
2N$ matrix
\begin{displaymath}
  \frac{\partial_l \partial_r S}{\partial \phi^A \partial \phi_B^*}\, ,
  ~~~A,B = 1, \, \ldots, N
\end{displaymath}
has rank $N$ on the stationary surface, defined by the equations of motion 
$\displaystyle{ \frac{\partial S}{\partial \phi} = \frac{\partial S}{\partial
  \phi^*} =0}$.
Notice that there is a natural grading among the 
fields and the anti-fields, namely the antighost number
\cite{gomis}. This allows a convenient  
decomposition of the extended action and, accordingly, the master
equation can be easily solved. \\
Because of the boundary conditions, a solution of the master equation
does not only give the dynamics of the  
system, as it generates the equations of motion; it also encodes the
gauge structure, as it generates the BRST transformations on a
function $X$ of fields and antifields via 
\begin{displaymath}
  s X = (X,S) \, .
\end{displaymath}
The master equation implies at once that $S$ itself is invariant under
this transformation, and that $s$ is a nilpotent differential: $s^2 =
0$. The second boundary condition ensures that $s$ ``starts as'' the
classical BRST differential, when acting on fields, but it is
automatically equipped with the modifications that are necessary when
the classical BRST transformations do not define a nilpotent operator;
this is the case of an open algebra, where the commutator of two gauge
symmetries
yields a gauge symmetry only on-shell. \\
Finally, the properness condition tells us that the extended action
has $N$ on-shell independent gauge symmetries, and this is just the
number that is necessary to eliminate unphysical anti-fields and
quantize only the fields. This is essential, since in the end one
wants to perform a path integral only on the space of fields
configurations; still, the antifields do not fall completely out of
the scene at this point: they can play the role
of classical sources for the BRST transformations. \\
It is useful, in the analysis of the cohomology of the BRST operator,
to introduce another grading, called anti-field number, whose value is
0 for the fields, 1 for the anti-fields of classical fields, 2 for
anti-fields of ghosts, 3 for anti-fields of ghosts for ghosts and so on.
The BRST operator can be decomposed accordingly:
\begin{displaymath}
  s = \delta + \gamma + \sum_{k \geq 0} s^{(k)} 
\end{displaymath}
The terms in the extended action all have non-negative antifield
number, and the antibracket operation lowers it by one, so that the
decomposition starts with degree -1, and the corresponding operator
$\delta$ which is necessarily nilpotent is called the Koszul-Tate
differential. It acts trivially on fields, while its action on an
anti-field yields the equation of motion of the corresponding field:
\begin{displaymath}
  \delta \phi_A^* = - \frac{\partial_l S}{\partial \phi^A} \, .
\end{displaymath}
The next term, $\gamma$, has antifield number 0. Its action on fields
is given by
\begin{displaymath} \label{eq:BRST}
  \gamma \phi^A = (\phi^A, S) |_{\phi^* = 0} \, .
\end{displaymath}

Canonical transformations are an important part of the formalism  
\cite{cano}. 
They preserve the anti-bracket structure (\ref{abracket}): 
calculating the anti-brackets in the old or
new variables is the same, or in other words the new variables also
satisfy (\ref{eq:brackffs}). Therefore also
the master equation $(S,S)=0$ is preserved. 

Canonical transformations from $\{\phi\, , \phi ^*\}$ to $\{\phi'\, ,
\phi'^*\}$,  
for which the matrix $\left. \frac{\partial_r \phi^B}{\partial
    \phi'^A}\right|_{\phi'^*}$ 
is invertible, can be obtained from a fermionic generating 
function $F(\phi,\phi'^*)$ with $gh(F) = -1$. The transformations are
defined by 
\begin{equation} \phi'^A=\frac{\partial_l F(\phi , \phi'^*)
}{\partial \phi'^*_A} \, , \hspace{2cm}
\phi^*_A=\frac{\partial_r F(\phi , \phi'^*)}{\partial \phi ^A}\,.
\label{Fcan}
\end{equation}
Infinitesimal canonical transformations are produced by 
\begin{displaymath}
  F = \phi '^*_A\phi ^A + \varepsilon f(\phi,\phi'^*) \, ;
\end{displaymath}
the first term is the identity operator in the space of
fields-antifields. 
In the following, we will have to use different types of transformations. 
\begin{itemize}
\item Point transformations are the easiest ones. These are just
fields redefinitions $\phi '^A=f^A(\phi )$. They are generated 
by $F=\phi '^*_A f^A(\phi )$ which thus determines the corresponding
transformations of the anti-fields. The latter replace the calculations of
the variations of the new variables.

\item Adding the BRST transformation of a function $s\Psi(\phi )$ to the
action is obtained by a canonical transformation with $F=\phi '^*_A\phi ^A+
\Psi (\phi )$. The latter gives
\begin{equation}
\phi'^A = \phi^A \ ;\qquad
\phi ^*_A =\phi'^*_A + \partial _A \Psi (\phi ).
\label{gfermion}\end{equation}
and by means of these transformations it is possible to implement the
gauge fixing  
conditions as a canonical change of variables. 

\item It is possible to redefine the symmetries by adding equations of motion
('trivial symmetries'). This is obtained by
\begin{equation}\label{symm_remov}
F=\phi '^*_A\phi ^A+ \phi '^*_A\phi '^*_B h^{AB}(\phi ) \, .
\end{equation}
\end{itemize}
The canonical transformations leave by definition the master equation
invariant, and because they are non-singular, they also preserve the
properness requirement on the extended action. This point will be
discussed at length in the following. Of course, in the new variables,
we do not see the classical limit anymore. But the most important
property is that the anti-bracket cohomology \cite{anti_coho} (and the
BRST cohomology) is not changed.

%%%%%%%%%%%%%%%%%%%%%%%%%%%%%%%%%%%%%%%%%%%%%%%%

\section{Strategy} 
\label{sec:strategy}

Before entering into the details of the analysis of the BS
superparticle~\cite{Brink-Schwarz}, we illustrate the main steps of
the quantization procedure of systems with an infinitely reducible gauge
symmetry. In particular, we describe each single canonical
redefinition of the field variables needed to bring the action in a
form which is suitable for path integral quantization. The main point
is to find out a correct gauge fixing procedure which fixes the gauge
degrees of freedom, and to construct the physical subspace as the
characteristic cohomology of the BRST operator.  In the present
section, we denote by $\Phi$ the ``physical'' variables
$X_\mu,\theta,e$ and $P_\mu$ (in the first order formalism), by
$k^\alpha_n$ the ghosts of the $\kappa$-symmetry and by $C_A$ the
ghost for the remaining local and rigid symmetries. We refer to the
index $n$ as level; the zero level is associated to the fields $\Phi^A$.

According to \cite{Lindstrom_BS,kallosh,GH,Tonin,Lindstrom}, the 
general solution of the master equation in the case of the BS
superparticle and of the GS superstring has the form
\begin{eqnarray}\label{eq:stru}
S &=& I[\Phi^A] + \Phi^{*}_A \left[ A^{A,B} ( \Phi) \, C_B +
  B^A_{~\alpha} ( \Phi)\,  k^{\alpha}_{1}  \right]  
\nonumber \\&+& C^{*}_A \left[ D^{A,B}(\Phi,C) \, C_B + 
F^A_{~\alpha\,\beta}(\Phi)\,  k^{\alpha}_{1}\,  k^{\beta}_{1} \right]
\nonumber \\&+&  k^{*,\alpha}_{n} \left[ G_{\alpha,\beta}(\Phi) \,
  k^\beta_{n+1} +  
F_{\alpha\,\beta}(C) \,  k^{\beta}_{n} + \hat{\delta}\, k^{\alpha}_{n}
\right]  
 \\&+& \Phi^{*}_A \Phi^{*}_B \left[ H^{A\, B}_{~\alpha \beta } ( \Phi)
   \,k^\alpha_{1}  k^\beta_{1}    
+ J^{A\, B}_{~\alpha} ( \Phi)\,  k^{\alpha}_{2}  \right] 
\nonumber \\&+&  C^{*}_A  \Phi^{*}_B   \left[ M^{A\,
    B}_{~\alpha\,\beta}(\Phi)\,  k^{\alpha}_{1}\,  k^{\beta}_{2}
\right] 
\nonumber \\&+&  k^{*,\alpha}_{n}  \Phi^{*}_B  \left[
  N^B_{\alpha,\beta}(\Phi) \, k^\beta_{n+2} +  
P^B_{\alpha, \beta\, \sigma}(C) \, k^{\alpha}_{1}  k^{\sigma}_{n+1} +
\hat{\hat{\delta}}\, k^{\alpha}_{n} \right] \,,\nonumber 
\end{eqnarray}
where the terms quadratic in the anti-fields are necessary to close
the algebra and to take into account the fact that  
the symmetry is reducible only on-shell, and $\hat\delta k_n,
~\hat{\hat\delta} k_n$ are of higher order in the ghosts.  
In the case of the BS
superparticle there are  further simplifications:  
{\it i)} the terms $F_{\alpha\,\beta}(C) \,  k^{\beta}_{n}$,
$\hat{\delta}\, k^{\alpha}_{n}$, and  
$ \hat{\hat{\delta}}\, k^{\alpha}_{n} $ are absent since the ghosts
are diffeomorphic invariant and they have no quartic  
interactions,  {\it ii)} the terms with powers of the anti-fields are
restricted to one type, that is when one of the anti-fields $\Phi^*$  
happens to be the anti-field of the einbein $e$. 

\begin{enumerate}
\item 
We begin the quantization procedure by introducing the ``twisting'' of the 
ghost fields. The aim of this redefinition is to decouple the ghost
fields $k^\alpha_n$ from the physical  
degrees of freedom described by  $X_\mu, P_\mu$ and $\theta$. For that
purpose, we will  
follow the ideas of P.~Tonwsend~\cite{3lectures} introducing redefined
ghost fields $\tilde{k}_n$  
and showing that their dynamics is independent of the ``physical'' variables. 
In particular, the idea is to redefine the fields in such a way that 
$\tilde{k}^{*,\alpha}_{n}  G'_{\alpha,\beta} \, \tilde{k}^\beta_{n+1}$ 
becomes independent of the field $\Phi$. Consequently, provided that
the gauge fixing  
procedure does not introduce a dependence on $\Phi^A$, the ghost
fields are decoupled from $\Phi^A$. 

As it will be explained in the next section, in order to implement the
suggested redefinition of ghost fields~\cite{3lectures}, we adopt a
twistor-like (off-shell) formalism. Obviously, this change of
variables can be performed by a canonical transformation. Here, we have
to point out that, as recognized by \cite{Bastianelli,kallosh}, some
canonical transformations involving infinite number of fields are
ill-defined and, therefore, they might change drastically the
BRST cohomology. It will be our main concern to avoid any such 
redefinition.
\item 
Furthermore, although the ``twisting'' provides a convenient
  coordinatization for the ghost degrees of freedom, there are
  remaining couplings between ghost fields and $\Phi^A$. After the
  ``twisting'', we have
\begin{eqnarray}
  \label{red.0}
   k^{*,\alpha}_{n} G_{\alpha \beta}(\Phi) \, k^\beta_{n+1} \longrightarrow  
  \tilde{k}^{*,\alpha}_{n} \left[ G'_{\alpha \beta} + 
    G^{'',A}_{ \alpha \beta} {\delta I \over \delta \Phi^A }\right]\,
  \tilde{k}^\beta_{n+1}\,.  
\end{eqnarray}
The second term in the bracket can be eliminated by a further
canonical transformation of type (\ref{symm_remov})
\begin{eqnarray}
  \label{red.1}
  \Xi = \Phi '^*_A\Phi ^A   + \sum_n   \tilde{k}^{'*,\alpha}_{n}
  \tilde{k}^\alpha_{n}   
+ \Phi '^*_A \sum_n \left( \tilde{k}^{*,\alpha}_{n} G^{'',A}_{\alpha
    \beta} \tilde{k}^\beta_{n+1} \right)\,.  
\end{eqnarray}
In the present case (and in the case of GS superstring),
$G^{'',A}_{\alpha \beta} {\delta I \over \delta \Phi^A }$ is
proportional to the equations of motion for the world-line (-sheet)
metric $e$ and, therefore, the canonical transformation $\Xi$ involves
the anti-field $e^*$. The latter is an anticommuting field and, thus,
$(e^*)^2 =0$. Consequently, it is easy to show that the redefinition
of the fields effected by $\Xi$ (and the corresponding inverse
transformation) involves only a finite number of fields $k_n$.
Obviously, the master equation and the classical cohomology of the
theory remain unchanged by these transformations and we can thus proceed
with the quantization procedure.

\item Following the BV formalism~\cite{BV}, one introduces a non-minimal
  sector of fields, that is the anti-ghosts, the extra-ghosts and the
  Lagrange multipliers
\begin{eqnarray}
  \label{antighosts}
  \bar C^\sigma &=& \{ \chi^q_p, \bar c_A \}\,, ~~~~~~
  \bar C^{*,\sigma} = \{ \chi^{*,q}_p, \bar c^{*}_A \}\,, \nonumber \\
  \pi^\sigma &=& \{ \pi^q_p, \pi^A \}\,,~~~~q + p = n, n \geq 1\,,
\end{eqnarray}
and the non-minimal part to the action
\begin{eqnarray}
  \label{non-minimal}
  S_{n.m.} = \bar c^{*}_A \, \pi^A + \sum_{p=0,q=1}^{\infty,q=p}
  \bar\chi^{*,q}_p \, \pi^q_p \,. 
\end{eqnarray}
However, since the application of the common BV formalism for an
infinitely reducible theory seems to generate wrong results
\cite{kallosh,Bastianelli}, we follow a different procedure. 

\item The main problem of a complete gauge fixing lies in the fact 
that after having fixed the minimal and extra ghosts, 
the theory is still invariant under a local symmetry of St\"uckelberg type 
\cite{stu}. The latter needs a further gauge fixing.  
As will be shown in Sec.~(\ref{sec:gauge}), some of the 
extra fields can be identified with the ghosts of  St\"uckelberg symmetry 
and, therefore, they can be used to fix completely the gauge invariances. 
As a consequence, the non-minimal sector of ghosts 
and extra ghosts is partially redundant and it can be reduced to a 
smaller set. Moreover, the gauge fixing of the entire ghost sector 
is performed in a step-by-step procedure 
by using canonical transformations which leave the 
cohomology unchanged.

\end{enumerate}

%%%%%%%%%%%%%%%%%%%%%%%%%%%%%%%%%%%%%%%%%%%%%%%%%%%%%

\section{The minimal solution of the Master Equation}
\label{sec:solution}

The general strategy is here applied to the specific model of 
the BS superparticle.  
The analysis of the classical action, the constraints coming from
the general coordinates invariance and the
supersymmetry on the target space  imply that the action is invariant
also under the $\kappa$-symmetry.  
We associate to  the world-line diffeomorphism the local ghost  
$c$, and  to the fermionic local $\kappa$-symmetry 
the infinite tower of ghosts $k_{n}$. 
This fixes the field content to (without considering the global ghosts
for the  super-Poincar\'e invariance)  
\begin{eqnarray}
  \label{eq:no_2}
  \{ \Phi^A \} &=& \{ e, X^\mu,  \theta^\alpha \}\,, ~~~~~~ {\rm gh}\,
  \Phi^A = 0  \,\nonumber \\ 
  \{ C^A \} &=& \{ c \}\,, \hspace{2.2cm} {\rm gh}\, C^A = +1 \nonumber \\
  \{ K^A \} &=&  \{ k_{n} \} \,, \hspace{2cm} {\rm gh}\, k_n = n \,,
\end{eqnarray}
and the corresponding antifields
\begin{eqnarray}
  \label{eq:no_3}
  \{ \Phi^{*}_A \} &=& \{ e^{*}, X^{*,\mu}, \theta^{*,\alpha} \}\,,
  ~~~~~~ {\rm gh}\, \Phi^*_A = - 1 , \nonumber \\ 
  \{ C^{*}_A \} &=& \{ c^{*} \}\,, \hspace{2.7cm}  {\rm gh}\, C^*_A = - 2
  \,, \nonumber \\ 
   \{ K^*_A \} &=&  \{ k^{*}_{n} \}\,,  \hspace{2.65cm} {\rm gh}\, k^*_n =
   - n -1 \,. 
\end{eqnarray}
 The classical action for BS superparticle \cite{Brink-Schwarz} is
 given, in first order formalism, by 
 \begin{equation}
   S = P^\mu \partial X_\mu - \bar\theta \SL{P} \partial\theta -{1
     \over 2}\, e 
   P^2 ; 
 \end{equation}
it is invariant under world-line reparametrizations, target-space
Poincar\'e transformations and the kappa-symmetry transformations 
\begin{eqnarray}
\delta X^\mu & = & \bar \theta \Gamma^\mu \SL{P} \kappa\,, \\
\delta \theta& = & \SL{P} \kappa\,, \nonumber \\
\delta e     & = & 4 \partial\bar\theta \kappa \,,\nonumber 
\end{eqnarray}
where $\kappa$, the parameter of the transformation, is an
anticommuting MW spinor, and is replaced by a commuting spinor ghost
$k$ in the BRST version of the symmetry.  The $\kappa$-symmetry is
closed on-shell (see e.g.\cite{green3}) and one can see easily that it
is reducible. The transformation has 16 parameters, the number of
components of a MW spinor in 10 dimensions, but on-shell they are not
all independent, since for $\kappa = \SL{P} \kappa'$ the symmetry is
trivial due to the equation of motion $P^2=0$. Thus only half of the
components of $\kappa$ can be used to gauge away degrees of freedom of
$\theta$, and on-shell one achieves the matching between bosonic and
fermionic degrees of freedom that is required by supersymmetry. The
other components are zero-modes of the symmetry.  There is no way to
get rid of these if one wants to keep the manifest Lorentz-invariance
of the model; instead, one has to consider this additional redundancy
as arising from a gauge symmetry in the ghost sector, namely
\begin{equation}
  \delta k = \SL{P} k_2 \, ,
\end{equation}
where an additional ghost has been introduced, which is a
MW spinor with chirality opposite to that of $k$.
The relation between $k$ and $k_2$ is the same as the one between
$\theta$ and $k$, so the new symmetry is again reducible, a third 
generation ghost is required, and so on: the symmetry is infinitely
reducible. 
According to the BV procedure \cite{BV}, the minimal solution of 
the master equation takes the form of an action depending on an
infinite number of fields. The matching of degrees of freedom is
obtained in this way: 8 components of $\theta$ are physical, the
remaining 8 are canceled by half of the components of the first ghost, 
the remaining components of $k$ are canceled by half of those of $k_2$ 
and so on. This can be written as
\begin{displaymath}
  8 = 8 + (8-8) - (8 - 8) + \ldots
\end{displaymath}
and clearly the sum makes sense only if it is performed with the
parenthesis in a certain position, or if it is regularized in a proper
way; it is customary to use the Euler regularization of an alternate
sum:
\begin{equation}
  \sum_{n=0}^\infty (-1)^n = \lim_{x\to 1} \sum_{n=0}^\infty (-x)^n =
  \frac{1}{2} \, .
\end{equation}
This gives a hint that the theory must be handled very 
carefully to avoid problems arising from formal manipulations of
non-convergent series. This kind of problems invalidate 
solutions given so far to the problem of covariantly quantizing the GS
string and the BS superparticle, as was first pointed out by Bastianelli 
et al. \cite{Bastianelli}.

The minimal solution of the master equation is \cite{Lindstrom_BS}
\begin{eqnarray}
 L & = & P^\mu \partial X_\mu - \bar\theta \SL{P} \partial\theta -{1
   \over 2}\, e 
   P^2 \nonumber \\
 && + X_\mu ^* (c P^\mu + \bar \theta \Gamma^\mu \SL{P} k_1 ) + e^*
 (\partial c + 4 \partial \bar \theta k_1 ) \nonumber \\
&& -2 c^* \bar k_1 \SL{P} k_1 + \bar\theta^* \SL{P} k_1 + \sum_{p \geq 1} 
\bar k_p^* \SL{P} k_{p+1} \nonumber \\
&& + 2 e^* \left[ X_\mu^*(\bar\theta \Gamma^\mu k_2 - \bar k_1
  \Gamma^\mu k_1) -4 c^* \bar k_1 k_2 + \bar \theta^* k_2 + \sum_{p
    \geq 1} \bar k_p^* k_{p+2} \right] 
\end{eqnarray}
This action satisfies the requirements requested to a minimal
solution. As expected, it is quadratic in the antifields since the
gauge algebra is open, but there are no terms with higher powers of
antifields.  

%%%%%%%%%%%%%%%%%%%%%%%%%%%%%%%%%%%%%%%%%%%%%%%%%%

\section{Twisting the ghosts}
\label{sec:twist}

The ghosts' zero mode condition, as we have seen, is field-dependent
and is only satisfied on-shell. These are the two unpleasant features
that we would like to eliminate, in order to get a well-defined
quantization procedure. Our strategy is to look for a redefinition of
the ghosts such that the new ghosts have off-shell zero modes, after
possibly some further manipulations that turn out to be necessary to
eliminate terms proportional to the equations of motion.  All the
transformations are performed via canonical transformations, so that
the master equation is satisfied at every step.  The first observation
to be made is that a twistor-like formulation\cite{twistor-like} is
possible for the massless particle or superparticle, in which bosonic
degrees of freedom are traded for world-line spinors, and --for
example in Galperin {\it et al.}-- the kappa-symmetry is replaced by
an extended world-line supersymmetry.  We adopt here a formulation
that differs from the previous ones in that the twistor-like variables
are not regarded as fundamental, but just as (non-linear)
functions of the fundamental fields. \\
The idea of the twisting is to replace, in the $\kappa$-symmetry
transformation laws, the operator $\SL{P}$ with another operator whose
nilpotency is independent of the equations of motion. The simplest
choice is to take a $2 \times 2$ matrix, and the only nilpotent
matrices are the Pauli matrices $\sigma_-$ and $\sigma_+$, up to a
multiplicative factor.  The fundamental equations that define the
``twistors'' $\lambda$ are the following:
\begin{eqnarray}
  \label{eq:twisting}
  (\SL{P})_\alpha^{~\beta} &=& \lambda_{\alpha,\underline{\alpha}\, i}
  (\sigma_- + P^2 
  \sigma_+)^{\underline{\alpha}}_{~\underline{\beta}} ~\delta^i_j
  ~\bar\lambda^{~\underline{\beta}j,\beta} \,
    \nonumber \\ 
  \bar\lambda^{~\underline{\beta}j,\beta}
  ~\lambda_{\beta,\underline{\alpha}i}  &=& 
   \delta^{\underline\beta}_{~\underline\alpha} ~\delta^j_i 
\end{eqnarray}
where we have defined $\bar \lambda = C_2^{-1} \lambda^T C_{10}$. As 
a consequence of the second equation of (\ref{eq:twisting}), the two- 
and ten-dimensional charge-conjugation matrices are related:
\begin{equation}
  C_2 = \lambda^T C_{10} \lambda \, .
\end{equation}
On shell, $\SL{P}$ is simply replaced by $\sigma_-$; the choice of
$\sigma_+$ is of course possible, and is related to the other choice
by a discrete symmetry, namely $\lambda \to \lambda C_2$.  From the
ten-dimensional point of view, the twistors $\lambda$ are commuting
spinors (it is worthwhile mentioning that although $\SL{P}$ is
real, the twistor can be complex), but they carry also internal
indices $\bar\alpha = 1,2 \, ; \, i= 1 , \ldots, 8$ in order to be
invertible matrices, as required by the second equation in
(\ref{eq:twisting}). We will never
need to display explicitly the whole index structure. \\
Due to the anticommuting properties of the $\sigma$ matrices, one can
verify that if eq.~ (\ref{eq:twisting}) is read as the definition
of a matrix $\SL{P}$, that matrix satisfies $\SL{P}^2 = P^2$.  But we
need to show that the equation can be solved, with respect to
$\lambda$, for any $P$. It is clear from the outset that the solution
can never be unique: for instance, $\lambda$ can be multiplied by an
orthogonal matrix $O_i^j$ in the internal indices. It can be proved
that these are the only internal transformations that leave
eq.~(\ref{eq:twisting}) invariant. By a rotation,
we can bring $P$ to have only the $P^+$ and $P^-$ components in
light-cone coordinates.  Then, using the well-known iterative
construction of $\Gamma$ matrices starting from the two-dimensional
ones, we have
\begin{eqnarray*}
\Gamma^+ = \left( \begin{array}{cc} 0 & 1 \\ 0 & 0 \end{array}
\right) \otimes  {\bf 1} \, , ~~~~
\Gamma^- = \left( \begin{array}{cc} 0 & 0 \\ 1 & 0 \end{array}
\right) \otimes  {\bf 1}
\end{eqnarray*} 
In this frame, $\Gamma^+$ and $\Gamma^-$ can be identified with $\sigma_+$
and $\sigma_-$.  \\
On-shell, $\SL{P}$ is a nilpotent matrix. We can choose a reference
frame in which $P^+$ is the only non-vanishing component of $P$; in  
this frame, $\SL{P}$ is proportional to $\Gamma^+$, whose kernel has a
dimension equal to one half of the dimension of the whole space.
The Jordan canonical form of $\SL{P}$ is a matrix with diagonal
$2\times 2$ blocks 
of the form of $\sigma_-$, and eventually a block of zeroes, but the
argument on the dimension of the kernel forbids the latter.
Then $\SL{P}$ can be transformed with a change of basis into
$\sigma_- \otimes 1 $, and this amounts to say that a 
solution to the twisting equation does exist. \\
Off-shell,  $\SL{P}$ is diagonalizable with eigenvalues $\pm
\sqrt{P^2}$. Since $\SL{P}$ is a linear combination of
$\Gamma$ matrices, $\textrm{tr} \SL{P} =0$. 
Then its Jordan form is  
\begin{equation} 
\sqrt{P^2} \left( \begin{array}{cc} 1 &0 \\ 0 & -1 \end{array} \right)
\end{equation}
which is the same Jordan form of $\sigma_- + P^2~ \sigma_+$, so 
we conclude again that eq.~(\ref{eq:twisting}) can be solved. \\
We have to stress that, even though the twistor is a non linear function 
of the momentum $P$, the Lorentz symmetry is linearly realized in the target space. 
The presence of an internal structure carried by the twisted ghosts does not affect the 
10-dimensional covariance. In addition, all degrees of freedom are 
quantized in a covariant manner, and the physical spectrum is identified 
with the antibracket cohomology.  In appearance, our formalism could remind the reader 
of a light-cone formulation of the superparticle, however the choice of $\sigma_\pm$ 
has been adopetd only for the internal indices $\underline{\alpha}$ of the twistors 
$\lambda^{\alpha, \underline{\alpha} i}$ and it is harmless for the covariance in the 
target space. 

After the twisting, the minimal solution has the form
\begin{eqnarray}
 L_{\rm min} & = & P^\mu \partial X_\mu - \bar\theta \SL{P} \partial\theta -{1
   \over 2}\,  
   P^2 (e -2 \sum \tilde {\bar k}_p^* \sigma_+ \tilde k_{p+1})  \nonumber \\
 && + X_\mu ^* (c P^\mu + \bar \theta \Gamma^\mu \SL{P} \lambda \tilde 
 k_1 ) + e^*
 (\partial c + 4 \partial \bar \theta \lambda \tilde k_1 ) \nonumber \\
&& -2 c^* \tilde {\bar k}_1 \bar\lambda \SL{P} \lambda \tilde k_1 +
\bar\theta^* \SL{P} \lambda \tilde k_1 + \sum  \tilde {
   \bar k}_p^* \sigma_- \tilde k_{p+1} \nonumber \\
&& + 2 e^* \left[ X_\mu^*(\bar\theta \Gamma^\mu \lambda \tilde k_2 -
  \bar {\tilde k}_1 \bar\lambda 
  \Gamma^\mu \tilde k_1) -4 c^* \tilde {\bar k}_1 \tilde k_2 + \bar
  \theta^* \lambda \tilde k_2 + \sum \tilde{\bar k}_p^* \tilde k_{p+2} \right]
\end{eqnarray}

The twisted ghosts have off-shell zero modes, given by $\delta \tilde
k = \sigma_- \tilde k'$ (from now on the tilde shall be understood,
and all the ghosts are twisted). These zero-modes coincide on-shell with the
original ones. We can now  eliminate completely the coupling of the
non-zero mode part of the ghosts, given by the terms $P^2 \bar
  k \sigma_+ k$, simply by a redefinition of the einbein
field. This is performed by means of a canonical transformation with
generating function
\begin{equation}
  F =  - 2 {e^*}' \sum_{p \geq 1} \bar {k_p^*}' \sigma_+ k_{p+1} \, . 
\end{equation}
The trivial part of the generating function, corresponding to the
identity operator, has been understood, as will be henceforth.
The corresponding shifts of the fields are listed below:
\begin{eqnarray}
  e'    &=& e - 2 \sum_{p \geq 1} \bar {k_p^*}' \sigma_+ k_{p+1} \nonumber\\
  k_p'  &=& k_p -2 e^* (-)^p \sigma_+ k_{p+1}  \\
  \bar k_{p+1}^* &=& {\bar k}_{p+1}^{*'} -2e^* \bar {k_p^*}' \sigma_+
  \nonumber 
\end{eqnarray}
These redefinitions as well as their inverses do not involve an
infinite number of fields due to the presence of the nilpotent factor
$e^*$, so they are allowed transformation in the sense discussed in
\cite{kallosh}, that is, the redefined field are not subject to any
constraint.
As a partial check of this statement, we shall now verify that the
BRST cohomology is not affected by the transformation.
The action is
\begin{eqnarray} \label{minaction}
 L_{\rm min}  & = & P^\mu \partial X_\mu -{1 \over 2}\, e P^2 -\bar\theta \SL{P}
 \partial\theta \nonumber \\
 && + X_\mu ^* (c P^\mu + \bar \theta \Gamma^\mu \SL{P} \lambda  k_1 ) + e^*
 (\partial c + 4 \partial \bar \theta \lambda k_1 ) -2 c^* \bar k_1
 \bar\lambda \SL{P} \lambda k_1  \nonumber \\
&& + \bar\theta^* \SL{P} \lambda k_1 + \sum 
   \bar k_p^* \sigma_- k_{p+1}\\
&& + 2 e^* X_\mu^*(\bar\theta \Gamma^\mu \lambda k_2 -
  \bar k_1 \bar\lambda 
  \Gamma^\mu \lambda k_1) -8 e^* c^* \bar k_1 \sigma_+ \sigma_- k_2 
 + 2 e^* \bar\theta^* \lambda \sigma_+ \sigma_- k_2 \, ; \nonumber 
\end{eqnarray}
One can notice that, as was to be expected, the terms $e^* k_n^*$ have
disappeared. 

The physical sector of the theory is given by the zero-ghost number
characteristic cohomology, that is, the cohomology of the BRST operator 
modulo the equations of motion. More precisely, two classical physical
observables are identified if they coincide on the equations of
motion; the identification is enforced by considering the cohomology
of the Koszul-Tate differential at antifield number 0, i.e. $H_0(\delta)$.
The BRST operator $\gamma$ defined by (\ref{eq:BRST}) anticommutes
with $\delta$ so that it defines a nilpotent operator in the
cohomology of $\delta$, and one can consider the group
$H(\gamma,H_0(\delta))$. This group is actually isomorphic to $H(s)$   
\cite{barnich}. \\
The relevant BRST transformations are
\begin{eqnarray}
 \gamma P &=& 0 \nonumber \\
 \gamma X &=& cP + \bar\theta \Gamma \SL{P} \lambda k_1 \nonumber \\
 \gamma \theta &=& \SL{P} \lambda k_1 \nonumber \\
 \gamma e &=& \partial c + 4 \partial \bar\theta \lambda k_1 
\end{eqnarray}
Manifestly $P$ is in the cohomology, and on-shell is a
constant light-like vector. It is easy to verify that the vector 
\begin{displaymath}
  U^\mu = (X \cdot P) X^\mu - {1\over 2} X^2 P^\mu - {1 \over 2}
  (\bar\theta \Gamma^{\alpha\beta\mu} \theta)X_{[\alpha}P_{\beta]}
\end{displaymath}
is BRST invariant, is light-like on-shell, and vanishes for $X$
proportional to $P$, so it accounts for 8 bosonic degrees of
freedom. The fermionic variables in the cohomology are $\SL{P}
\theta$.

%%%%%%%%%%%%%%%%%%%%%%%%%%%%%%%%%%%%%%%%%%%%%%%%%

\section{Gauge-fixing}
\label{sec:gauge}

We have now written the minimal extended action, i.e. a minimal proper
solution of the master equation, in a form that is suitable for
further manipulations.  The next task to be performed is the
gauge-fixing of the $\kappa$-symmetry. This requires the introduction
of a set of non-minimal fields and non-minimal terms in the extended
action. We shall follow a completely standard procedure: the
non-minimal fields include extra-ghosts $\chi$ and Lagrange
multipliers $\pi$, with opposite statistics, and the non-minimal terms
have generically the form $\bar\chi^* \pi$.  We shall work by steps,
at each step fixing the gauge of a given level of fields.  At level
$0$ (the classical fields), we have
\begin{equation}
  L_0 = L_{\rm min} + L_{\rm nm,0} = L_{\rm min} + {\bar\chi}_1^{1*} \pi_1^1 \, ;
\end{equation}
the gauge-fixing of $\theta$ is achieved by the gauge fermion
\begin{equation}
  \Psi_0 = \bar\chi_1^1 \, \partial(\bar\lambda \theta) \,;
\end{equation}
the canonical transformation generates new terms in the action:
\begin{equation}
  L_0 \to L_0 + \bar\theta \, \lambda \partial\pi_1^1 -
  \partial\bar\chi_1^1 \bar\lambda \, \SL{P} \lambda
  k_1 - 2 e^* \partial \bar\chi_1^1 \, \sigma_+ \sigma_- k_2 ; 
\end{equation}
and the term proportional to the equations of motion which appears
after rewriting $\SL{P}$ in the twisted form can be eliminated by the
canonical transformation
\begin{equation}
  \Theta = 2 {e^*}' \partial \bar\chi_1^1 \sigma_+ k_1
\end{equation}
that gives
\begin{equation}
  L_0' =   L_0 + \bar\theta \, \lambda \partial\pi_1^1 -
  \partial\bar\chi_1^1 \sigma_-
  k_1 + 2 e^* \bar k_1 \sigma_+ \partial \pi_1^1 \,  .
\end{equation}
The propagator for the physical field $\theta$ is now invertible, as
is required for a gauge-fixed action. \\
At level 1, 
\begin{eqnarray}
  L_{\rm nm,1} &=&  {\bar\chi}_2^{1*} \pi_2^1 \, , \nonumber \\
  \Psi_1 &=& \bar\chi_2^1 \partial k_1 \, ,  \\
  L_1'  &=& L_0' + ({\bar\chi}_2^{1*} - \partial \bar k_1) \pi_2^1 - \partial
  \bar\chi_2^1 \, \sigma_- k_2 \, ; \nonumber
\end{eqnarray}
we have fixed the gauge of $k_1$. 
A crucial remark is now in order \cite{kallosh}: the ghost
$\chi_1^1$ 
not only has the usual and expected $\kappa$-symmetry, but also a more
general one, of St\"uckelberg type, namely
\begin{eqnarray*}
  \delta \chi_1^1 =  \epsilon \, ,~~~~~~~
  \delta \pi_2^1 =  \sigma_- \epsilon \, . 
\end{eqnarray*}
The gauge-fixed action (the part independent of the antifields) is
invariant under this symmetry. We take into account this new symmetry
introducing a new ghost, $\omega_1$, and new terms in the
action. These have the form of minimal terms for the new symmetry, but 
it turns out that other terms are necessary to close the master
equation:
\begin{equation}
  L_{\rm St}^{(1)} = ({\bar\chi}_1^{1*}  + {\bar \pi}_2^{1*} \sigma_- 
  + {\bar\chi}_3^{3*}) \omega_1 - {\bar\chi}_2^{1*} \sigma_-
  \chi_3^3 \, .
\end{equation}
It can be verified that $L_1' + L_{\rm St}^{(1)}$ satisfies the master
equation, and the cohomology is not altered.
This is hardly surprising, when one realizes that $L_{\rm St}^{(1)}$ is
just a non-minimal term written in transformed variables. Indeed, the
non minimal term is
\begin{displaymath}
  {\bar\chi}_3^{3*} \omega_1
\end{displaymath}
and $L_{\rm St}^{(1)}$ is recovered via the transformation
\begin{displaymath}
  \Psi = ({\bar\chi}_1^{{1*}'} + {\bar\pi}_2^{{1*}'} \sigma_-) \chi_3^3 \, .
\end{displaymath}
With a canonical transformation we now redefine the new ghost
$\omega_1 \to \omega_1 - \pi_1^1$ ; as a result, the non-minimal term
${\bar\chi}_1^{1*} \pi_1^1$, previously introduced, is now canceled by the
shift of the first term in $ L_{\rm St}^{(1)}$. We can now proceed to fix 
the St\"uckelberg symmetry; although it allows for an algebraic
fixing, we choose to use a Lorentz gauge-fixing, in order to have more 
uniformity in the treatment of the various extra-ghosts.  We then add
further non-minimal terms
\begin{equation}
  L_{\rm nm,1}' = {\bar\chi}_2^{2*} \pi_2^2 + {\bar\chi}_3^{1*} \pi_3^1 \, 
  ;
\end{equation}
\begin{displaymath}
  \Psi_1' = \bar\chi_2^2 \partial \chi_1^1 + \bar\chi_3^1 \partial k_2 
  \, ;
\end{displaymath}
\begin{eqnarray}
  L_2  &= &  L_1' +  L_{\rm St}^{(1)}(\omega_1 \to \omega_1 - \pi_1^1)
  \nonumber \\ 
&+&  ({\bar\chi}_2^{2*} -\partial \bar\chi_1^1 ) \pi_2^2 - \partial
\bar\chi_2^2 \, \omega_1 \\
&+& ({\bar\chi}_3^{1*} -\partial \bar k_2 ) \pi_3^1
   - \partial \chi_3^1 \sigma_- k_3 \, . \nonumber
\end{eqnarray}
It is now apparent that the role of the ghost $\omega_1$ is that of a
Lagrange multiplier, which fixes the gauge of $\chi_2^2$; thus there
is no need to introduce other non-minimal terms and gauge-fixings for
$\chi_2^2$ that would bring in the action a mixing of $\chi_1^1$ with
higher-level ghosts. \\
The first step is completed; at this point, the ghost $k_2$ is fixed,
while $\chi_2^1$ has the St\"uckelberg symmetry. As before, we
introduce non-minimal terms ${\bar\chi}_4^{3*}~\omega_2$ followed by 
\begin{displaymath}
\Psi =
({\bar\chi}_2^{{1*}'} - {\bar\pi}_3^{{1*}'} \sigma_-) \chi_4^3 \, ,  
\end{displaymath}
that yields
\begin{equation}
L_{\rm St}^{(2)} = ({\bar\chi}_2^{1*}  - {\bar\pi}_3^{1*} \sigma_-  
+ {\bar\chi}_4^{3*}) \omega_2 + {\bar\chi}_3^{1*} \sigma_-
\chi_4^3 \, ;
\end{equation}
then we perform a canonical transformation 
\begin{eqnarray}
  \omega_2 & \to & \omega_2 - \pi_2^1 + \sigma_- \chi_3^3 \nonumber \\
  {\bar \pi}_2^{1*} & \to & {\bar \pi}_2^{1*} + {\bar\omega}_2^* \\
  {\bar\chi}_3^{3*} &\to & {\bar\chi}_3^{3*} - {\bar \omega}_2^* 
  \sigma_-  \, . \nonumber
\end{eqnarray}
It is actually sufficient to substitute in the action the shift of
$\omega_2$, since the shifts of the antifields do not generate any new
term, as they cancel each other. But this canonical transformation leads
also to the cancellation of the terms ${\bar\chi}_2^{1*} \pi_2^1$ and
${\bar\chi}_2^{1*} \sigma_- \chi_3^3$. The gauge-fixing is done in
perfect analogy with the previous step:
\begin{equation}
  L_{\rm nm,2} = {\bar\chi}_3^{2*} \pi_3^2 + {\bar\chi}_4^{1*} \pi_4^1 \, 
  ;
\end{equation}
\begin{displaymath}
  \Psi_2 = \bar\chi_2^2 \partial \chi_1^1 + \bar\chi_3^1 \partial k_2 
  \, ;
\end{displaymath}
\begin{eqnarray}
  L_3 &=& L_2 +  L_{\rm St}^{(2)}(\omega_2 \to \omega_2 - \pi_2^1 + \sigma_- 
  \chi_3^3) \nonumber \\
 &+&  ({\bar\chi}_3^{2*} - \partial \bar\chi_2^1 ) \pi_3^2 - \partial
 \bar\chi_3^2 \omega_2 \\ 
 &+& ({\bar\chi}_4^{1*} - \partial \bar k_3 ) \pi_4^1
   - \partial \bar\chi_4^1 \sigma_- k_4 \, . \nonumber
\end{eqnarray}
Now the ghosts $k_3, \chi_2^1$ and $\chi_3^2$ are fixed, while
$\chi_3^1$ has a St\"uckelberg symmetry. \\
We are now able to give the algorithmic procedure for obtaining the
gauge-fixed action at any level: given the action at level $n$, when the 
last ghost fixed is $k_n$ and $\chi_n^1$ has the St\"uckelberg
symmetry, one has to perform the following operations: \\
1) introduce the non-minimal terms corresponding to the
St\"uckelberg symmetry
\begin{equation}
  L_{\rm St}^{(n)} = ({\bar \chi}_n^{1*} + (-)^{n+1}
  {\bar\pi}_{n+1}^{1*} \sigma_- 
  + {\bar \chi}_{n+2}^{3*} ) \omega_n + (-)^n {\bar\chi}_{n+1}^{1*} \sigma_- 
  \chi_{n+2}^3 \, ;
\end{equation}
2) redefine the St\"uckelberg ghost:
\begin{equation}
  \omega_n \to \omega_n - \pi_n^1 + (-)^n \sigma_- \chi_{n+1}^3 \, ;
\end{equation}
this has to be done via a canonical transformation, so there are
also other redefinitions which, however, do not have any effect on the 
action; \\
3) add non-minimal terms
\begin{equation}
  L_{{\rm nm},n} = {\bar\chi}_{n+1}^{1*} \pi_{n+1}^1 + {\bar\chi}_{n+1}^{2*}
  \pi_{n+1}^2 \, ; 
\end{equation}
4) fix the gauge with the canonical transformation
\begin{equation}
  \Psi_n = \bar\chi_{n+1}^1 \partial k_{n+1} +  \bar\chi_{n+1}^2
  \partial \chi_n^1 \, . 
\end{equation}
The procedure just consists of canonical operations which can not
change the cohomology. 
The situation at level 3 is depicted in Table \ref{tbl:fields4}. 
We find in this way that only a subset of the whole pyramid of
extra-ghosts as is given e.g. in \cite{GH} is really necessary for
the quantization procedure: an entire sub-pyramid with
vertex in $\chi_3^3$ can be omitted. 
It is easy to count the degrees of freedom and verify that the
counting is correct at each step. 
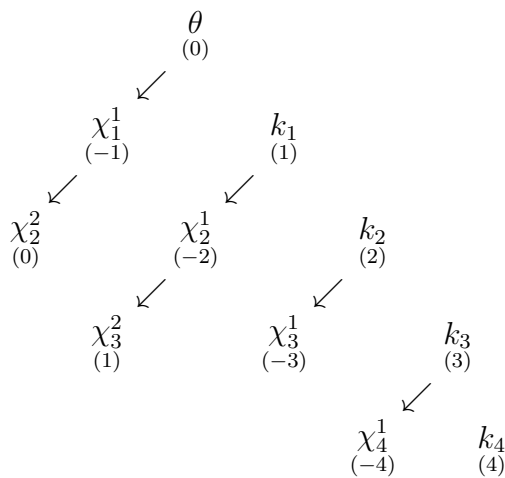
\tabcolsep 1pt
\begin{table}[thb]\caption{Fields after the 3rd step}
\label{tbl:fields4}\begin{center}\begin{tabular}{ccccccccccccccccc}
 &   &   &   &   &   &   & &\ghe{\theta}0 &   &   &   &   &   &   &
 &  \\
 &  &   &   &    &   &   & $\swarrow$ & &    &   &   &   &   &   &   &  \\
 &  &      &   &   &   & \ghe{\chi_1^1}{-1}  &   &   &
&\ghe{k_1}1 &   &
 &   &   &&   \\
 &  &      &   &   & $\swarrow$   &   &   &   & $\swarrow$   &   &   &   &
 &   &  &    \\
 &  &   &   &    \ghe{\chi_2^2}0  &
   &  && \ghe{\chi_2^1}{-2}  &
 &   &   &\ghe{k_2}2     &   &   &\\
 &  &   &    &   &   &   & $\swarrow$   &   &   &   & $\swarrow$
 &   &      & &   & \\
 &  &
  &
&   &   & \ghe{\chi_3^2}1
&  
 & &&\ghe{\chi_3^1}{-3}  &
 &   &   & &\ghe{k_3}3  &   \\

   & &   &   &   &    &   &   &   &    &
 & &   &$\swarrow$  &  &   &     \\
 &    & &
&   &   &
& 
& & &&&\ghe{\chi_4^1}{-4}  &
 &   &   & \ghe{k_4}4    \\
\end{tabular}\end{center}\end{table}     
\tabcolsep 6pt

It is useful to look at the Lagrangian at a generic step $n$. Its form 
is given by
\begin{eqnarray}
  \label{eq:action_n}
  L_n   &= & L_{\rm min} + 2 e^* \bar k_1 \sigma_+ \partial \pi_1^1 +
  \bar\theta  
  \lambda \, \partial \pi_1^1  \nonumber \\
 & - & \sum_{p=1}^{n+1}
 \partial \bar \chi_p^1 \, \sigma_- k_p - \sum_{p=1}^n \partial \bar k_p
 \, \pi_{p+1}^1  
 - \sum_{p=1}^{n-1} \partial \bar\chi_p^1 \, \pi_{p+1}^2 
 - \sum_{p=2}^n \partial \bar\chi_p^2 \, \omega_{p-1} \nonumber \\
 & + & \sum_{p=1}^{n-1} {\bar\chi}_p^{1*} \, \omega_p  
  + \sum_{p=2}^n {\bar\chi}_p^{2*} \, \pi_p^2 + \sum_{p=3}^n
 {\bar\chi}_p^{3*} ( \omega_{p-2} - \pi_{p-2}^1 + (-)^p \sigma_-
 \chi_{p-1}^3 ) \nonumber \\
 & + & \sum_{p=2}^n (-)^p {\bar\pi}_p^{1*} \sigma_- (\omega_{p-1} -
 \pi_{p-1}^1 )   \nonumber \\
 & + &  {\bar\chi}_n^{1*} 
 ( \pi_n^1 - \sigma_- \chi_{n+1}^3 ) + {\bar\chi}_{n+1}^{1*}
 \, \pi_{n+1}^1  \, . 
\end{eqnarray}

The complete action, fixed at every level of ghosts, is now obtained
simply as the formal limit $n \to \infty$: all the sums in
(\ref{eq:action_n}) run up to infinity, and the two terms in the last
line must be dropped.  The equations of motion fix all fields to
constant values.  

Finally, the invariance under diffeomorphisms is algebraically fixed 
by introducing the non-minimal fields (a Lagrangian multiplier $d$ and the 
antighost $b$), a non -minimal term in the action and performing the 
canonical transformation generated by $\Psi_{\rm diff}~$: 
\begin{equation}
  \label{diff_ga}
  L_{\rm diff} = b^* d \,, ~~~~~ \Psi_{\rm diff} = b \, ( e - 1)\,. 
\end{equation}
They form a trivial pair under BRST, namely $s \, b = d$ and $s \, d = 0$, so they 
do not belong to the cohomology.

%%%%%%%%%%%%%%%%%%%%%%%%%%%%%%%%%%%%%%%%%%%%%%%%%%%

\section{The BRST cohomology}
\label{sec:cohomology}

We will now compute the BRST cohomology for the model in the limit $n
\to \infty$. At each step of our procedure, it is guaranteed that the
cohomology is correct. The limit is the only operation which is not
canonical; nevertheless, we will show now that the cohomology is still
the correct one.  
The relevant parts of the BRST transformations for the fermionic 
sector, in the limit $n \to \infty$, are listed below:

\begin{eqnarray}  
\begin{array}{ll}
  s \,  \theta   =  \SL{P} \lambda k_1 \, + 2 b \lambda \sigma_+
    \sigma_- k_2 \,,  \hspace{1cm}      
   & s \,  \theta^*  =  \SL{P} \partial \theta + \lambda \partial
    \pi_1^1   \,,  \nonumber \vspace{.15cm}\\ 
  s  \,  k_p   =  \sigma_- k_{p+1} \,,   \hspace{1cm}    
   & s \,  k_p^*  = (-)^p \sigma_- \, ( \partial \chi_p^1 - 
    k_{p-1}^* )\, + \partial \pi_{p+1}^1 \,,  \nonumber
    \vspace{.15cm}\\ 
  s \,  \chi_p^1  =  \omega_p   \,,   \hspace{1cm} 
   & s \,  \chi_p^{1*}  =  \sigma_- \partial k_p + \partial
     \pi_{p+1}^2 \,, \nonumber \vspace{.15cm}\\ 
  s  \,  \chi_p^2  =  \pi_p^2  \,, \hspace{1cm}     
   & s \,  \chi_p^{2*}  =  \partial \, \omega_{p-1}  \,, \nonumber
    \vspace{.15cm}\\ 
  s  \,  \chi_p^3 =  \omega_{p-2} - \pi_{p-2}^1 + (-)^p \sigma_-
    \chi_{p-1}^3  \,, \hspace{1cm}      
   & s \,  \chi_p^{3*}  = \sigma_- \chi_{p+1}^{3*}  \,, \nonumber
    \vspace{.15cm}\\ 
  s  \,  \pi_p^1  = (-)^p \sigma_- (\omega_{p-1} - \pi_{p-1}^1 )  \,,
  \hspace{1cm}      
   & s \,  \pi_p^{1*}  = (-)^p \, (\partial k_{p-1} + \chi_{p+2}^{3*}
   ) \, + \sigma_- \, \pi_{p+1}^{1*}   
      \,, \nonumber \vspace{.15cm}\\ 
  s  \,  \pi^2_p  =  0   \,, \hspace{1cm}     
   & s \,  \pi_p^{2*}  =  (-)^p \, ( \partial \chi_{p-1}^1 - \chi_p^{2*} ) \,,
    \nonumber \vspace{.15cm}\\ 
  s  \,  \omega_p  =  0  \,, \hspace{1cm}     
   & s \,  \omega_p^*  =  (-)^p \, ( \partial \chi_{p+1}^2 -
   \chi_p^{1*} - 
    \chi_{p+2}^{3*} ) - \sigma_- \pi_{p+1}^{1*}  \,, 
\end{array} 
 \end{eqnarray}
 \\
 where we have omitted the terms coming from the part of the action
 quadratic in the antifield in eq. (\ref{minaction}), as well as the
 terms dependent from the antifields of the bosonic sector and of the
 diffeomorphism sector. These terms do not change the results of the
 analysis.  In the above list of transformations, one can recognize at
 once some trivial pairs, e.g. $(\chi_p^1, \omega_p)$, $(\chi_p^2,
 \pi_p^2)$; then one can define a reduced cohomology, setting to zero
 the trivial pairs in the above transformations.  Splitting the spinor
 fields in two components, $\kappa = (\kappa^+, \kappa^-)$, so that
 $\sigma_- \kappa = (\kappa^-, 0)$, one can see that also $(k_p^+,
 k_{p+1}^-)$, $(\pi_{p}^{1-}, \pi_{p+1}^{1+})$ are trivial pairs, as
 well as $(\chi_{p}^{3-}, \chi_{p+1}^{3+})$, and finally
 $(\pi_{1}^{1+},\chi_{3}^{3+})$. We conclude that the only field in the
 physical spectrum is $\SL{P} \theta$. Moreover, a careful analysis of
 the antifield sector along the same lines shows that $\SL{P} \partial
 \theta$ is actually BRST-exact, so that only its zero mode survives
 in the cohomology. In this way, we recover the correct superparticle
 spectrum.
 
 In the limit, we obtain an action that is free, {\it i.e.} quadratic
 in all fields, except for non-polynomial interactions involving the
 twistor-like fields and cubic interactions involving only the
 physical fields, the diffeomorphism ghost , and the first
 $\kappa$-symmetry ghost $k_1$, as well as the corresponding Lagrange
 multipliers. All the other ghosts are decoupled, so they do not enter
 in the computation of any amplitude involving only external physical
 states.

\section{Conclusions}

Using the conventional BV-BRST methods, we provide the complete
gauge-fixed action for the Casalbuoni-Brink-Schwarz superparticle.
Despite several difficulties, by means of a suitable redefinition of
the minimal sector fields in terms of twisted variables, we are able
to construct a quantization procedure of the model. A direct
computation of the antibracket cohomology shows that the BRST charge
selects the correct physical spectrum of the theory. In addition, it
is also verified that the characteristic cohomology
$H(\gamma,H(\delta))$ confirms the result of the antibracket analysis.

\section*{Acknowledgements}
\noindent
We thank Peter van Nieuwenhuizen and Warren Siegel for illuminating
comments and suggestions. The research has been supported by the NSF
grants no. PHY-9722083 and PHY-0070787.

%%%%%%%%%%%%%%%%%%%%%%%%%%%%%%%%%%%%%%%%%%%%%%%%%%

\appendix

\section{Notations and conventions}
\label{app:notations}

Our conventions are the following.  $X^m$, $m = 0, 1,\ldots, 9$,
denotes the 10d 
space-time coordinates and $\Gamma^m$ are 10-d Dirac matrices 
\begin{equation}
\{\Gamma^m, \Gamma^n\} = 2\eta^{mn}, \quad {\rm where}~ \eta = (- ++ \ldots +).
\end{equation}
These $\Gamma$'s differ by a factor of $i$ from those of
ref.~\cite{green3}.  For this choice of gamma matrices the massive
Dirac equation is $(\Gamma\cdot\partial - M)\Psi =0$.  In fact, our
conventions are such that the quantity $i=\sqrt{-1}$ will not appear
in any equations. As it is quite standard, we also introduce
$\Gamma_{11} = \Gamma_{0}\Gamma_{1}\ldots \Gamma_{9}$, which satisfies
$\{ \Gamma_{11}, \Gamma^m \} =0$ and $\big(\Gamma_{11}\big)^2 =1$ and
the totally antisymmetrized Dirac matrices $\Gamma^{m n \dots k} =
\Gamma^{[m}\Gamma^{n}\ldots \Gamma^{k]}$.

The Grassmann coordinates $\theta^\alpha$ are space-time spinors 
and world-line scalars. They can be decomposed as
$\theta = \theta_R + \theta_L$, where
\begin{equation}
\theta_R = {1 \over 2} (1 + \Gamma_{11})\,\theta, \qquad 
\theta_L = {1 \over 2}(1 - \Gamma_{11})\,\theta. \label{thetas}
\end{equation}
For anticommuting spinors, the  rule to lower and to rise the spinor
indices implies 
\begin{eqnarray}
  \label{ch_1}
&& \xi^\alpha = C^{\alpha \beta} \xi_\beta\,, \nonumber \\
&& \frac{\partial}{\partial\theta^\alpha} = C_{\beta\alpha}
\frac{\partial}{\partial\theta_\beta}\,, \\ 
&& \bar\theta \chi = \theta^\alpha \chi_\alpha = - \theta_\alpha
\chi^\alpha =  \chi^\alpha  \theta_\alpha = \bar\chi \theta\,, 
\nonumber 
\end{eqnarray}
where $ C^{\alpha \beta}$ is the antisymmetric charge conjugation
matrix.  With these conventions, the fermionic degrees of freedom of
the BS superparticle are carried by a Majorana-Weyl (MW) spinor
$\theta_R$. In the same way
all the $\kappa$-symmetry ghosts are MW spinors. \\
In general, the symmetry properties of bilinears built out of MW
spinors are the following \cite{bergshoeff}:
\begin{equation}
  \bar\psi \Gamma_{\alpha_1 \ldots \alpha_n} \chi =
  (-1)^{1+\epsilon(\psi)\epsilon(\chi)+n(n+1)/2} ~\bar\chi
  \Gamma_{\alpha_1 \ldots \alpha_n} \psi .
\end{equation}
There is only one independent non-vanishing bilinear of a single
MW-spinor, i.e. $\bar\chi
\Gamma_{\alpha \beta \gamma} \chi$.

%%%%%%%%%%%%%%%%%%%%%%%%%%%%%%%%%%%%%%%%%%%%%%%%%%%%%

%%%%%%%%%%%%%%%%%%%%%%%%%%%%%%%%%%%%%%%%%%%%%%%%%%%%

\end{document}